\documentclass[12pt,a4paper]{article}

\usepackage{graphics,graphicx}
\usepackage{epsfig}
\makeatletter
\@addtoreset{equation}{section}
\makeatother

\begin{document}
\vspace{5cm}
\begin{flushright}
QMUL-PH-05-08\\
hep-th/0504226
\end{flushright}
\vspace{1cm}
\begin{center}
\Large \textbf{Inflation from Geometrical Tachyons}\\
\vspace{2cm}
\normalsize \textbf{Steven Thomas} \footnote{s.thomas@qmul.ac.uk} 
 \textbf{and John Ward}\footnote{j.ward@qmul.ac.uk}\\
\vspace{2cm}
\normalsize 
\emph{ Department of Physics \\
Queen Mary, University of London \\
Mile End Road, London \\ E1 4NS  U.K \\}
\end{center}
\vspace{2cm}
\begin{center}
\textbf{Abstract}
\end{center}
We propose an alternative formulation of tachyon inflation using the
geometrical tachyon arising from the time dependent motion of a BPS $D3$-brane
in the background geometry due to  $k$ parallel $NS$5-branes arranged 
around a ring of radius $R $. Due to the fact that the 
mass of this geometrical tachyon field is  $\sqrt{2/k} $ times smaller than 
the corresponding open-string tachyon mass, we 
find that the slow roll conditions for inflation and the number
of e-foldings can be satisfied in a manner that is consistent with an 
effective 4-dimensional model and with a perturbative string coupling.  
We also show that the metric
perturbations produced at the end of inflation can be sufficiently small and do
not lead to the inconsistencies that plague the open string tachyon models.
Finally we argue for the existence of a minimum of the geometrical tachyon 
potential which could give rise to a traditional reheating mechanism.

\newpage
\section{Introduction.}
Despite making many promising advances toward a complete theory of quantum gravity, string/M theory has perhaps not made  similar progress in the area of inflationary cosmology where there is an abundance of 4D theoretical models and experimental data \cite{carroll, linde}.
The main focus has effectively  been on two types of models. The first is where there is some kind
of spontaneous compactification of the extra dimensions of spacetime, leaving us with our familiar
four dimensional universe and the standard model. The second model
is that the universe itself is located on the world volume of a 3-brane perhaps in 
a higher dimensional bulk, with the standard model living on the world-volume.
Both approaches have proven to be illuminating, yet the problem of inflation
in string theory models still proves elusive.

One of the simplest proposals for inflation in a string theory context, has been that
of tachyonic inflation \cite{gibbons}, where the tachyon plays the role of the inflaton.
In the past few years there has been great progress in understanding the nature
and role of tachyons in string theory \cite{sen}, in particular it has emerged that many
of the features of the tachyon can be captured surprisingly well by an effective
DBI action \cite{tachyon_action}. Tachyons arise in several contexts in the theory, most notably
in the latter stages of brane-antibrane annihilation \cite{carroll} but also as open string degrees of freedom
on a Non-BPS brane \cite{sen}. This has stimulated several papers on tachyon inflation and cosmology
 \cite{cline, piao, joris, li, fairbairn}.

However there are several problems associated with open-string tachyon cosmology \cite{kofman2, shiu}
which appears to cast doubt over the tachyon's role in inflation. 
The first, and most serious, is that the potential for the field is a runaway exponential, tending to
its asymptotic minimum at $T=\pm \infty$. 
\footnote{Note however that in the case of non-BPS branes in superstring theory, there is evidence that
the tachyon potential for static fields develops localized minima at finite values of $T$, see e.g. \cite{sen2}}
 
Thus, not only is this far too steep to generate the required number of e-foldings but there is no minimum for the tachyon to oscillate around and generate reheating \cite{kofman1, sami}. The second problem is that at the start of inflation
the de Sitter radius of the universe is actually smaller than the string length and thus an
effective theoretical description breaks down, a consequence of this is that the string coupling is
large in this region and so perturbative analysis cannot be used.
There is also the additional problem
of the tachyonic energy, which dominates during inflation and therefore dominates for all time,
although arguably this is not as problematic as the other two objections.
Whilst there have been several ingenious attempts to bypass these problems, most notably 
\cite{cline, joris, li, bento}, it seems unlikely that the open-string tachyon could be responsible for inflation, although it may still play a role as dark matter fluid \cite{sen} or
as part of a pre-inflationary phase \cite{kofman1, linde}.

Recent work on time dependent solutions in the linear dilaton background of coincident $NS$5-branes,
has shed new light on a possible geometrical description of the open string tachyon \cite{time_dependence, kutasov, thomas}. In particular,
it is conjectured that radion fields on a probe $Dp$-brane become tachyonic when the probe moves in a bounded, compact space.
In addition, the mass scale of this geometrical tachyonic mode is substantially smaller than that of the open string tachyon, and therefore may resolve some
of the problems associated with inflation.

In this note we will examine the effective action for a $D3$-brane with a geometrical tachyon on its world volume, described by a cosine potential \cite{thomas}. The cosine potential arises naturally in
the context of 'Natural Inflation' \cite{freese} due to the creation of pseudo Nambu-Goldstone bosons arising from symmetry breaking, whereas in our
case the cosine potential arises due to the geometry of the background brane configuration. It is tempting to identify parameters in the geometrical
picture with that of Natural Inflation, however the non-linear 
form of the tachyon effective action makes this difficult.
We will see how our theory fits in with the
inflationary paradigm, and whether conventional reheating is possible. We begin by reviewing the
construction of the geometrical tachyon solution and the basics of tachyon
cosmology. We will then check the consistency of the theory with regards to the slow roll
and e-folding approximations before attempting to calculate various perturbation amplitudes and provide a
discussion of reheating. We close with some remarks and possible implications for future work.
\section{String Background.}
We begin this note with a brief description of the string theory background
associated with our model. We will consider the $NS$5-brane background of type II string
theory, where we have $k$ parallel but static fivebranes localised on
a circle of radius $R$. The branes will be assumed to be unresolvable for the moment, which
can be interpreted as a smearing of the charge of $k$ branes around a ring configuration.
The background solutions for these fivebranes are given by the CHS solutions \cite{CHS}
\begin{eqnarray}\label{eq:ring}
ds^2 &=& \eta_{\mu \nu} dx^{\mu} dx^{\nu} + H(x^n)dx^m dx^m \\
e^{2(\phi-\phi_{0})} &=& H(x^n) \nonumber \\
H_{mnp} &=& -\varepsilon^q_{mnp} \partial_q \phi \nonumber,
\end{eqnarray}
where $\mu, \nu $ parameterize the directions parallel to the fivebranes and $m, n$ are
the transverse directions and $\phi$ is the dilaton. $H(x^n)$ is the harmonic function which describes the
orientation of the fivebranes in the transverse space, which in our simplistic case
\footnote{The exact form of the harmonic potential for $k$ $NS$5 branes arranged around a ring,
 as computed in \cite{sfetsos}, is rather more complicated than the expression in (\ref{eq:ring2}).
 The latter form emerges as 
an approximation valid for distances $ r >> {2\pi R}/k $ i.e. effectively a large 
$k$ approximation. In this
limit the $NS$5-branes appear as smeared around the ring. \cite{thomas}}
is given by \cite{sfetsos}
\begin{equation}\label{eq:ring2}
H = 1+ \frac{kl_s^2}{\sqrt{(R^2+\rho^2 + \sigma^2)^2-4R^2\rho^2}}.
\end{equation}
In the above solution we have switched to polar coordinates
\begin{eqnarray}
x^6 &=& \rho \cos(\theta), \hspace{1cm} x^7=\rho \sin(\theta) \\
x^8 &=& \sigma \cos(\psi), \hspace{0.9cm} x^9 =\sigma\sin(\psi), \nonumber
\end{eqnarray}
and now the harmonic function describes a ring oriented in the $x^6-x^7$ plane and
 we have an $SO(2) \times SO(2)$ symmetry in the transverse space. In our analysis, $l_s$ is the string length and we will be working entirely in string frame. Now the above ring configuration is supersymmetric in 10D, however the introduction
of a probe $Dp$-brane breaks this supersymmetry entirely and the probe brane will
experience a gravitational force pulling it toward the fivebranes. To avoid further complications
we will always assume that $3 \le p \le 5$, since for $p <3$ there is a divergence due
to the emission of closed string modes which will render the classical theory
useless \cite{sahakyan}. Note that we are also able to switch between IIA and IIB theory because the $NS$5 brane
background in insensitive to T-duality, as the harmonic function couples only to the transverse parts of the
metric. We now insert a probe $Dp$-brane in this background whose low energy effective action is the DBI action,
which we write as
\begin{equation}
S=-\tau_{p} \int d^{p+1} \zeta e^{-(\phi-\phi_0)} \sqrt{-det(G_{\mu \nu}+B_{\mu \nu}+\lambda F_{\mu\nu})},
\end{equation}
where both $G_{\mu \nu}$ and $B_{\mu \nu}$ are the pullbacks of the space-time tensors
to the brane, $\lambda = 2\pi l_s^2$ is the usual coupling for the open string modes, $F_{\mu \nu}$ is the field strength of
the $U(1)$ gauge field on the world volume, and $\tau_p$ is the tension of the brane.
We will assume that the transverse scalars are time dependent only, and set the
gauge field and Kalb-Ramond field to zero for simplicity. Upon insertion of the
$NS$5-brane background solution, we see the action in static gauge reduces to the simple form
\begin{equation}
S=-\tau_p \int d^{p+1} \zeta \sqrt{H^{-1}-\dot{X}^2},
\end{equation}
with $X^m$ parameterizing the transverse scalar fields.
In order to find our geometrical tachyon, we consider motion of the probe brane in the
 plane of the ring ($ i.e. \, \sigma = 0 $ and in the interior of the ring and map the above action to a form that is familiar from the non-BPS action for  open-string tachyons. The result is that we have \cite{tachyon_action, kutasov, thomas}
\begin{equation}\label{eq:gtachyon}
S= -\int d^{p+1}\zeta V(T) \sqrt{1-\dot{T}^2},
\end{equation}
where the potential is given by
\begin{eqnarray}\label{eq:cosine}
V(T)&=& \tau_p^u \cos\left(\frac{T}{\sqrt{kl_s^2}}\right) \hspace{0.2cm} \\
\tau_p^u &=& \frac{\tau_p R}{\sqrt{kl_s^2}}, \nonumber
\end{eqnarray}
and the tachyon can be expressed as a function of the coordinate $\rho$ as
\begin{equation}
T(\rho) = \sqrt{kl_s^2} \arcsin(\rho/R).
\end{equation}
In obtaining the above, we have used the throat approximation for the harmonic function, which means 
neglecting the factor of unity in (\ref{eq:ring2}). This may not be necessary, but it does allow an exact expression to be obtained. 
Under this assumption, we see that taking $\rho = \sigma=0$ in (\ref{eq:ring2}) (i.e. the centre of the ring) 
requires that $ \sqrt{k} l_s >> R$, which is the first constraint we find on the parameters $k, l_s $ and $R$.
Later on we will use numerical techniques to arrive at a form of the potential 
$V(T)$ which will use the exact form of the harmonic function as calculated in \cite{sfetsos}. In principle
we can then relax the throat approximation which
leads to the cosine potential in (\ref{eq:cosine}) so that the previous inequality may not be needed.

To avoid confusion with the open-string tachyon, from now on (unless otherwise stated)
 we will use tachyon to refer to the geometrical tachyon in (\ref{eq:gtachyon}).

We see that the tachyon potential is symmetric about the origin, which arises as a consequence of the background geometry.
It should be noted that this mapping is non-trivial in the sense that we began by probing a gravitational
background, and have ended up with a solution in flat Minkowski space. This tells us that
there are two equivalent ways of visualizing the theory.
Firstly there is the bulk viewpoint, where there
is actually a ring of $NS$5-branes and the solitary probe brane universe moving in the throat geometry.
Alternatively, we could view the problem as a single brane moving in flat space-time with
a highly non-trivial field condensing on its world volume.
In what follows we will find it useful to switch between these two pictures in order to
better understand the physics.
In fact the bulk viewpoint is even more complicated as we know \cite{kutasov, thomas}
that the tachyon field has a geometrical interpretation as a BPS brane in a confined, bounded
space, but we could also describe it as a non-BPS brane which has
a soliton kink stretched across the interior of the ring \cite{thomas}.

Clearly we see that the geometrical tachyon varies between $T=\pm \pi \sqrt{kl_s^2}/2$
in contrast to the usual open string tachyon which is valued between
$\pm \infty$. This is due to the probe brane being confined \emph{inside} the
ring. Expanding the potential about the unstable vacuum yields a tachyonic mass of
$M_T^2 = -1/kl_s^2$. For sufficiently large $k$ this can be made much smaller than the open string tachyonic mass
$M^2=-1/2$ (in units where the string length is unity). It is this different mass scale and
profile of the potential which suggest that the geometrical tachyon may be useful in describing
inflation on a $D3$-brane.
The energy momentum tensor can be calculated in the usual way, and has non-zero components
\begin{eqnarray}
T_{00} &=& \frac{V(T)}{\sqrt{1-\dot{T}^2}} \\
T_{ij} &=& -\delta_{ij} V(T) \sqrt{1-\dot{T}^2} \nonumber
\end{eqnarray}
from which we see that the pressure of the tachyon fluid tends to
zero as the tachyon rolls toward the zero of $ V(T)$.

In \cite{thomas} we found that there was also probe brane motion through the ring in the $x^8-x^9$ plane.
Although this did not lead to the creation of a geometrical tachyon, one could imagine
a scenario where the probe oscillates in this direction through the ring, radiating energy
as it does so. Eventually the probe would settle at the origin, which is an unstable point
in the ring plane corresponding to $T=0$, and we then recover our geometrical tachyon solution.
\section{Tachyon Cosmology.}
In a cosmological context, the condensing tachyon will generate a gravitational field
on the probe $D3$-brane and therefore we must include this minimal coupling in the action.
We will also assume that there is no coupling to any other string mode in order to keep the
analysis simple, however we should be aware that there is no reason why other modes should
not be included \cite{choudhury}.
Our Lagrangian density can thus be written
\begin{equation}
\mathcal{L} = \sqrt{-g} \left( \frac{R}{16\pi G}-V(T)
\sqrt{1+g^{\mu\nu}\partial_{\mu}T \partial_{\nu} T }\right),
\end{equation}
where $g^{\mu \nu}$ is the metric and $R$ is the usual scalar curvature. For
simplicity we will assume that there is a FLRW metric of the form
\begin{equation}
ds^2= -dt^2+a(t)^2 dx_i^2,
\end{equation}
with $i$ running over the spatial directions. We have implicitly assumed here that we have a flat universe, which is acceptable because any curvature
is negligible in the very early stages of the universe.
The effect of the scale factor is to modify the energy density, $u$ for the
flat background such that it
is no longer conserved in time \cite{gibbons}, instead we find
\begin{equation}
u=\frac{a^3 V(T)}{\sqrt{1-\dot{T}^2}},
\end{equation}
which prevents us from obtaining an exact solution for the tachyon in the presence of the
gravitational field in the usual manner.
From this we can determine the late time behaviour of the tachyon condensate. If we assume that
$u$ is constant, then the pressure will vary as $p = -V(T)^2 u$ and will tend
to zero as $V(T)$ reaches its minimum. Using the standard equation of state $p=\omega u$, we find
that $\omega =-(1-\dot{T}^2)$ which implies $-1 \le \omega \le 0$.
From the Lagrangian density, we can also obtain the Friedman and Raychaudhuri equations
for the tachyon condensate
\begin{eqnarray}
H^2 &=& \left(\frac{\dot{a}}{a}\right)^2= \frac{\kappa^2 V(T)}{3 \sqrt{1-\dot{T}^2}} \\
\frac{\ddot{a}}{a} &=& \frac{\kappa^2 V(T)}{3\sqrt{1-\dot{T}^2}} \left(1-\frac{3\dot{T}^2}{2} \right),
\end{eqnarray}
where $\kappa^2 = 8\pi G = M_{p}^{-2}$, $M_p = 2.2\times 10^{18} GeV$ and the cosmological constant term
is set to zero. There is also a  useful relationship between the 4D Planck mass and the
string scale obtained via dimensional reduction \cite{jones}
\begin{equation}\label{eq:reduction}
M_p^2 = \frac{v M_s^2}{g_s^2},
\end{equation}
where $M_s = l_s^{-1}$ is the fundamental string scale and the quantity $v$ is given by
\begin{equation}
v = \frac{(M_s r)^d}{\pi},
\end{equation}
with $r, d$ being the radius and number of compactified dimensions respectively (typically $d$=6 for the superstring).
Note that for our effective theory to hold we require $v >> 1$.
Obviously we are assuming there is some unknown mechanism which stabilizes
the compactification manifold, and freezes the moduli so that they do
not interfere with our tachyon solution.

The evolution of the universe is effectively determined by the Raychaudhuri equation which
shows that inflation will cease when $\dot{T}^2 = 2/3$ and the universe will then decelerate as
the tachyon velocity increases.
Upon variation of the action, we find the equation of motion for the tachyon field can be written
\begin{equation}\label{eq:eom}
\frac{V(T) \ddot{T}}{1-\dot{T}^2}+3HV(T)\dot{T} + V'(T) = 0,
\end{equation}
where a prime denotes differentiation with respect to $T$, and $H$ is the Hubble parameter.
Note that in deriving this equation we must also use the conservation of entropy of the tachyon fluid
\cite{gibbons}.
We see that $3HV(T)\dot{T}$ acts as a friction term, in much the same way as in standard inflationary
models, except that this term may vanish for the open string tachyon as the field rolls to $\pm \infty$ where its potential vanishes.
For a scalar field to be a candidate for the inflaton it must satisfy the usual slow roll parameters
as well as providing enough e-foldings during rolling. The tachyon is no exception \cite{steer}, and so we use the
conventions employed in \cite{li, hwang} to write the slow roll parameters
\begin{eqnarray}
\epsilon(T) &=& \frac{2}{3}\left(\frac{H'(T)}{H^2(T)} \right)^2 \\
\eta(T) &=& \frac{1}{3} \left(\frac{H''(T)}{H^3(T)} \right). \nonumber
\end{eqnarray}
Where we assume that the acceleration of the tachyon is negligible, and require that $\epsilon <<1$ and $|\eta|<<1$ in order to generate inflation.
There appears to be little or no consensus on the correct slow roll equations to use for the tachyon,
however we see that the same general behaviour is obtained even if we use the conventions in
\cite{piao} or \cite{fairbairn}.
The number of e-folds produced between $T_o$ and $T_e$ is given by
\begin{equation}\label{eq:efoldings}
N(T_o, T_e) = \int_{T_o}^{T_e} dT \frac{H}{\dot{T}},
\end{equation}
which must satisfy $N \sim 60$ to agree with observational data \cite{carroll}. $T_o$ and $T_e$ are two
arbitrary points on the potential where the slow roll conditions are still satisfied.
\section{Geometrical tachyon inflation.}
One of the many problems associated with tachyon inflation is that the mass scale of the
open string tachyon is simply too large. However we have seen that this is not necessarily the case
where we have a geometrical tachyon, and so we may enquire whether inflation is possible in this
instance.
We first consider the geometrical tachyon starting very close to the top of its potential with a small
initial velocity to ensure that it will roll. Note that if $T=0$ then spontaneous symmetry breaking will
cause the universe to fragment into small domains which will each have differing values of the tachyon field.
Inflation can occur only if $H^2 >> |M_T^2|$ near the top of the potential, which translates into the
constraint
\begin{equation}
\frac{\tau_3 R}{3 M_p^2} >> \frac{1}{\sqrt{kl_s^2}}.
\end{equation}
Where $\tau_3$ is the tension of a stable $D3$-brane.
Since we are considering the large $k$ limit, the RHS is very small and so we find that this condition is satisfied. \footnote{In the finite $k$ case we will have to be careful to ensure that this
constraint is fulfilled.}
Furthermore this also suggests that the effective theory for geometrical tachyons is valid because we can clearly see that
\begin{equation}
(kl_s^2)^{1/2} >> H^{-1},
\end{equation}
and so the de-Sitter horizon may be larger than the string length for large $k$. This is in contrast
to the open string scenario where we find that the horizon is smaller than the string length, and thus
should not be described by an effective theory.

In order to check that this is correct, we must try and get a handle on the size of the string coupling.
In the open string tachyon case we find that in order to satisfy $H^2>>|M_T^2|$ at the top of the potential,
we have the constraint
\begin{equation}
g_s >> 260 v.
\end{equation}
As already mentioned, the effective theory is only valid for $v>>1$, which implies that we are in the
strong coupling regime and therefore an effective theory may not yield reliable results.
In contrast, a similar calculation for the geometrical tachyon implies
\begin{equation}\label{eq:top}
g_s >> \frac{24 \pi^3 v }{\sqrt{k} R M_s} \approx  744 \frac{v}{\sqrt{k} R M_s},
\end{equation}
thus by fixing appropriate values for $k$ and $R$ we may ensure that $v>>1$ and also that $g_s << 1$. Earlier we saw the throat condition $ \sqrt{k}l_s >> R $
which means that $ \sqrt{k} R M_s << k $. Thus, in fact it is large values for $ k$ that will essentially allow a satisfactory solution to (\ref{eq:top}). For example, assuming $v = 10$ a value of $k\approx 10^5 $ would allow for perturbative $g_s$ to solve (\ref{eq:top}). Relaxing the throat approximation may allow much smaller values of $k$.
This is
interesting because we see that the weak coupling arises entirely due to a parameters describing the background brane solution in the bulk
picture. In contrast, there is not a clear explanation to account for the origin of the weak self-coupling of the
inflaton in standard single scalar field \cite{steer} inflation.

Despite this apparent success we may be concerned that the effective theory may still not be a valid description at the
top of the potential \cite{kofman2}. In order to check this we should compare the effective tension of the unstable brane to
the Planck scale. After some algebra, and using the equation for weak coupling we find
\begin{equation}
\frac{\tau_3^{u}}{M_p^4} \sim \frac{3}{k^2}\left(\frac{24 \pi^3}{R M_s} \right)^2 v.
\end{equation}
Again we see that for a certain range of background parameters (and assuming $R M_s >> 1$), the effective tension need not
be Super-Planckian and therefore the DBI can still be a good approximation to 4D gravity.

As there is an obvious similarity with Natural Inflation \cite{freese}
we could demand that the height of the potential to be of the order of $M_{GUT}^4$ (where $M_{GUT} \sim 10^{16}$ GeV)
in order to generate inflation, however we will try to keep this arbitrary for the moment.
Using the potential, we immediately see that the slow roll conditions for our geometrical tachyon
can be written as
\begin{eqnarray}
\epsilon &=& \frac{M_p^2}{2\tau_3 R \sqrt{kl_s^2}} \frac{\tan^2(T/\sqrt{kl_s^2})}{\cos(T/\sqrt{kl_s^2})}\\
\eta &=& \frac{-M_p^2}{4\tau_3 R \sqrt{kl_s^2}} \frac{\left(1+\cos^2(T/\sqrt{kl_s^2})\right)}{\cos^3(T/\sqrt{kl_s^2})}.
\end{eqnarray}
Slow roll will only be a valid approximation when the tachyon is near
the top of its potential and thus
we may effectively neglect the contribution from the trigonometric functions as they will only be terms of $\mathcal O(1)$.
In fact for small $T$ we see that $\epsilon$ is already extremely small.
Dropping all numerical factors of order unity gives us the primary constraint for slow roll;
\begin{equation}
M_p^2 << \tau_3^u R \sqrt{kl_s^2},
\end{equation}
which must be satisfied by both equations. Given that the reduced Planck mass in string theory is typically of the
order of $2.4 \times 10^{18}$ GeV, this means that $k$ and  $R/l_s$ must be large.
Generally the slow roll conditions will be satisfied due to the
mass scale of the geometrical tachyon, for large $k$. In the open string models, the larger
mass implies that the tachyon may only have been involved in some pre-inflationary phase. Of course,
the analysis in both cases is classical and quantum corrections may well prove to be important is determining the
exact behaviour near the top of the potential.

We can estimate the number of e-foldings using (\ref{eq:efoldings}), however this turns out to be sensitive to the
value of the tachyon velocity near the top of the potential. To remedy this we use the equations of motion and the
slow roll approximation, which allows us to re-write this equation in terms of
the potential and its derivative. In fact this is the method most commonly used in standard inflationary analysis.
After some algebra we obtain
\begin{eqnarray}\label{eq:efoldings2}
N_e &=& -3 \int dT \frac{H^2 V(T)}{V'(T)} \\
&=& \frac{\tau_3 R \sqrt{kl_s^2}}{M_p^2} \left\lbrace\cos\left(\frac{T_e}{\sqrt{kl_s^2}}\right)-
\cos\left(\frac{T_o}{\sqrt{kl_s^2}}\right)+\ln\left(\frac{\tan(T_e/2\sqrt{kl_s^2})}{\tan(T_o/2\sqrt{kl_s^2})}\right) \right\rbrace \nonumber
\end{eqnarray}
Using the constraint from the slow-roll equations we see that the leading term must be large. If we demand that $T_o$ and $T_e$ are reasonably close, then the contribution
from the other terms will be small, and so the number of e-foldings will depend on the ratio
\begin{equation}
\nu = \frac{\tau_3R \sqrt{kl_s^2}}{M_p^2},
\end{equation}
where $\nu \ge 60$ in order for there to be enough inflation. However if we dont impose this restriction, but allow inflation to begin near the
top of the potential and end near the bottom, then there can be significant contribution to the number of e-foldings from the
additional terms. This has the effect of reducing the value of $\nu$ - however it must still satisfy the slow roll constraint
of being larger that unity.
We can write the unstable brane tension in terms of the string coupling, string mass scale and the parameter $\nu$, thus we have  the height of the potential
given by
\begin{equation}
M_{infl}^4 = \frac{M_p^2 M_s^2 \nu}{k},
\end{equation}
which defines our effective inflation scale $M_{infl}$.
The exact value of $M_s$ depends on the particular string model but it is usually assumed to lie in the range
1 Tev - $10^{16}$ GeV. So as an example, if  $\nu \sim 60$,  $M_s \sim 10^{16}$ GeV
and $k \sim 10^5 $ we find $M_{infl} \sim 10^{16}$ GeV.
\subsection{Numerical Analysis.}
We can also check the consistency of our analytic solutions by numerically solving for the Hubble parameter.
We can write the Hubble equation as a function of $T$ rather than time (since the tachyon field is monotonic
with respect to time - at least initially), and then using the Friedmann equation we obtain the following
first order differential equation \cite{fairbairn}
\begin{equation}\label{eq:diff1}
H^{\prime 2}(T) - \frac{9}{4}H^4(T)+ \frac{1}{4M_p^4}V(T)^2=0,
\end{equation}
where a prime denotes differentiation with respect to $T$. Solving this for the Hubble term gives us
a constraint on the velocity of the tachyon field
\begin{equation}\label{eq:velocity}
\dot{T}^2 = 1 - \left(\frac{V(T)}{3M_p^2 H(T)^2} \right)^2.
\end{equation}
It will be convenient to work with dimensionless variables in our numerical analysis, so we
define the dimensionless tachyon field and Hubble parameter as follows,
\begin{equation}
y=\frac{T}{\sqrt{kl_s^2}}, \hspace{0.5cm} h(y)=\sqrt{kl_s^2}H(y).
\end{equation}
We can solve (\ref{eq:diff1}) to obtain $h(y)$ and then substitute this into (\ref{eq:velocity})
to determine the velocity of the tachyon field. We choose the initial velocity of the field to be
zero, and the initial value of $T_o$ to be very small. As in \cite{fairbairn}, the general behaviour is dependent upon
the dimensionless ratio $X_0$, where $X_0^2=\nu$.
Some results are plotted in Figure 1. We find that the velocity (strictly speaking this is the square of the velocity) of the tachyon field is very
small over a large range, only becoming large as it nears the bottom of the potential.
In inflationary terms this implies that universe will be inflating for almost the entire duration of the rolling of the field.
For increasing values of $X_0$, inflation ends at larger values of $T$. However even for the case of $X_0=2$, which barely
satisfies the slow roll constraints, we expect inflation to end reasonably close to the bottom of the potential.
\begin{figure}[ht]
\begin{center}
\epsfig{file=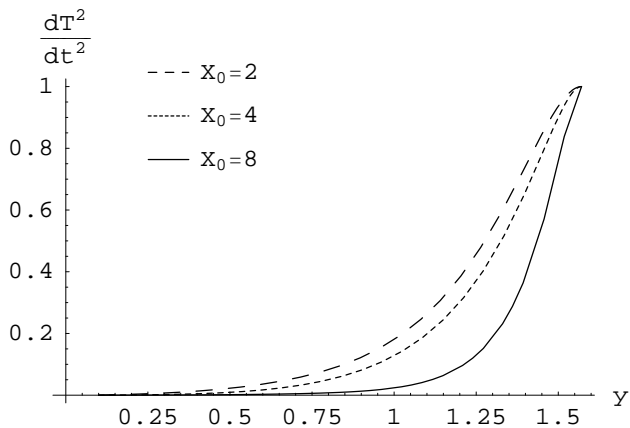, width=10cm,height=6cm}
\caption{Velocity of tachyon field for differing values of $X_0$, with an initial velocity of zero.}
\end{center}
\end{figure}
\begin{figure}[ht]
\begin{center}
\epsfig{file=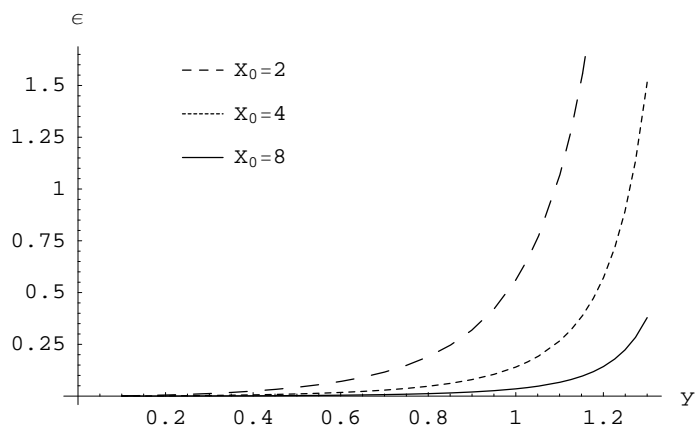,width=10cm,height=6cm}
\caption{Value of the slow-roll parameter $\epsilon$ for various
values of $X_0$}
\end{center}
\end{figure}
\begin{figure}[ht]
\begin{center}
\epsfig{file=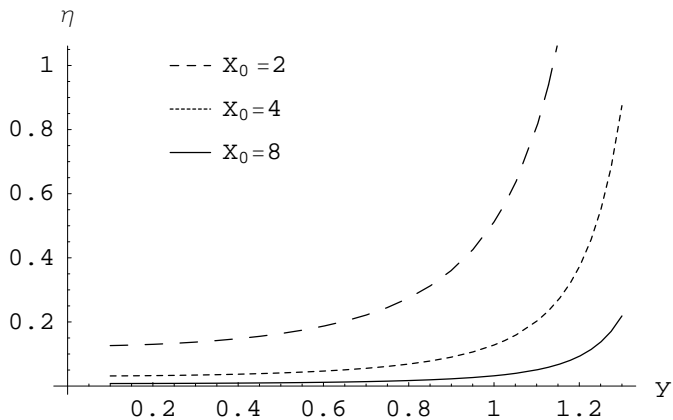,width=10cm,height=6cm}
\caption{Value of the slow-roll parameter $\vert {\eta }\vert $ for various values of
$X_0$}
\end{center}
\end{figure}
We can also make a numerical check on the smallness
of the slow roll parameters $\epsilon, \eta $ (see figs 2 and 3). Using our numerical
solution for $h$ we can also determine the amount of e-foldings during inflation
via (\ref{eq:efoldings2}). It turns out that in order to generate at least 60 e-foldings we only need to take
$\nu \sim 30$

Finally, we can also use numerical techniques to try and reconstruct the tachyon potential by 
using the full form of the ring harmonic function as derived in
 \cite{sfetsos} without assuming the approximation that lead to the cosine
potential (\ref{eq:cosine}). Recall that this approximation was that the $NS$5 branes were 
unresolvable as point sources arranged uniformly around the ring.
As our tachyon field rolls from near the top of the cosine potential down towards the value 
$T/{\sqrt{k}l_s} = \pi/2$, the geometric picture of this process is that we start from near 
the centre of the ring at $\rho = 0$ and move towards the ring located at $\rho = R $. As  
$T/{\sqrt{k}l_s}$ nears $\pi/2$, even for large $k$, the approximation of a continuous distribution
of $NS$5 branes around the ring will break down and individual sources will be resolvable. It is at
 this point that we expect the true potential $V(T)$ to deviate from the cosine form. Fig 4 shows
 the shape of the potential one obtains for the case $k=1000$, by numerically
implementing the tachyon map discussed earlier, using the exact form of the ring harmonic function.
In this plot we have chosen the angular variable $\theta$ that appears in the exact
form of the harmonic function to be fixed at $\pi/{2k} $ for simplicity. 
Details of the harmonic function 
relevant to fully resolvable $NS$5 branes are given later on in Section 5.
 What is perhaps most apparent about this potential is the existence of a minimum very close to the ring
 location. It also turns out that our previous cosine potential is an excellent approximation to this
  numerical plot
for values of $T $ to the left of the minimum. Later on in section 5, we shall see how analytic methods
 can be used to verify the existence of this minimum.
\begin{figure}
\begin{center}
\includegraphics[width=10cm,height=6cm]{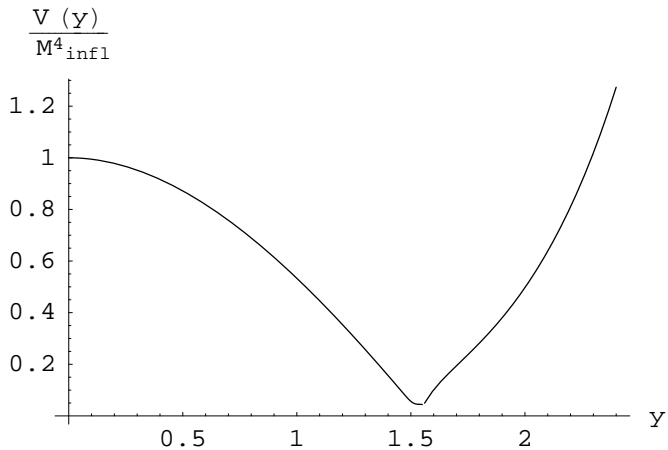}
\end{center}\caption{Profile of the tachyon potential taking $k=1000$. The solutions from each region are matched onto each other
at $T=\pi/2$ in dimensionless units.}
\end{figure}

\subsection{Perturbations.}
So far, so good, but we must also consider the perturbation fluctuations generated at the end of inflation.
One of the generic difficulties associated with open-string tachyonic inflation is the fact that the tension of
the $D3$-brane must be significantly larger than the Planck mass. This indicates that the effective action
cannot adequately describe 4D gravity, as it will have metric fluctuations that are always too large \cite{kofman2}.
This is not the case for our geometrical tachyon as we seen there are additional scales in the theory which can reduce the
overall effect of these fluctuations.
There are two main perturbations to consider, the scalar, and the gravitational (tensor) ones which we will denote by
$\mathcal{P_T}$ and $\mathcal{P_G}$ respectively. (Strictly speaking, $\mathcal{P}$ corresponds to the
amplitude of the perturbation). Constraints
from observational data imply the relation \footnote{Thanks to D. J. Mulryne for pointing this out.}
\begin{equation}
|\mathcal{P_T}| + |\mathcal{P_G}| \le 10^{-5}.
\end{equation}
During inflation, gravitational waves are produced whose amplitude is given by $\mathcal{P_G} \sim \frac{H}{M_p}$, but
observational data of the anisotropy of the CMB \cite{kofman2, li, fairbairn} implies that at the end of inflation
\begin{equation}\label{eq:tensor}
\frac{H_{end}}{M_p} \le 3.6 \times 10^{-5},
\end{equation}
and we must ensure that this condition is consistently satisfied in order for us to consider the geometrical
tachyon as a possible candidate for the inflaton.
In order to verify this we will first consider the scalar perturbation and use the solution from that to determine our
mass scales for the metric perturbations, as it is generally more important to see whether the tachyon action
allows for small metric fluctuations. For simplicity we will assume that inflation ends when the following
constraint is satisfied \cite{kofman2} 
\begin{equation}
H \sim |M_T| \sim \frac{M_s}{\sqrt{k}},
\end{equation}
and we will also assume that the tachyon velocity at this time is given by $\dot{T}=\sqrt{2/3}$.
The scalar perturbations are determined in the usual manner using
\begin{equation}
|\frac{\delta \rho}{\rho}| \sim \frac{H\delta T}{\dot{T}},
\end{equation}
where $\delta T$ satisfies the following constraint near the top of the potential \cite{fairbairn, parameters}
\begin{equation}
\delta T \sim \frac{H^2}{2\pi \sqrt{V(T)}}.
\end{equation}
Combining the last two equations we write the amplitude for the scalar perturbation as
\begin{equation}\label{eq:scalar}
\mathcal{P}_T \sim \frac{H^2}{2\pi \dot{T} \sqrt{V(T)}} \le 10^{-5}.
\end{equation}
(We should actually calculate the values of $H$ and $\dot{T}$ during inflation in order to determine the
ratio of the perturbations, however since we expect $T$
to be a slowly varying field (\ref{eq:scalar}) should remain constant over a large range of wavelengths \cite{linde}.)
If we assume that $T$ is small then the cosine part of the potential is close to unity, and upon substitution of the Hubble term we find
\begin{equation}\label{eq:scalar_result}
\mathcal{P}_T \sim \frac{M_s}{M_p}\sqrt{\frac{3}{8\pi^2k \nu}} \le 10^{-5},
\end{equation}
We can use this to determine a constraint upon the string scale/Planck scale ratio as follows
\begin{equation}\label{eq:ratio}
\frac{M_s}{M_p} \le \sqrt{\frac{8 \pi^2k\nu}{3}} \times 10^{-5}.
\end{equation}
As an example, for $k \sim 10^3$ and $\nu \sim 28$ (\ref{eq:ratio}) implies $ M_s  \leq 10^{16} GeV$

Solving the equation for the metric perturbation leaves us with
\begin{equation}
\mathcal{P}_{G} \sim \frac{H}{M_p} \sim \frac{M_s}{M_p \sqrt{k}} \le 3.6 \times 10^{-5}
\end{equation}
which is explicitly dependent upon this ratio. 
We can establish the absolute upper bound on the perturbation using (\ref{eq:ratio})
\begin{equation}
\mathcal{P}_{G} \le 2\pi \times 10^{-5} \sqrt{\frac{2\nu}{3}}.
\end{equation}
If $\nu$ is $\mathcal{O}(30)$ then this implies the maximum perturbation will
be of the order of $10^{-4}$ which is slightly too large. However in general we may expect
that the metric perturbations will be acceptably small by assuming a smaller string scale than the one that saturates (\ref{eq:ratio}) for given $k$. This is encouraging since the open string tachyon always admits
large metric fluctuations, and therefore cannot be responsible for the last 60 e-foldings of
inflation \cite{kofman1}. In our case these fluctuations can be suppressed when $k$ is sufficiently large,
and we can find inflationary behaviour leading to the correct amount of structure formation.
\section{Reheating.}
Perhaps the most problematic aspect of tachyon inflation is the shape of the potential itself. 
The open string tachyon potential is exponentially decaying at large field values with its minimum at asymptotic
infinity. Thus even if it were possible to satisfy all the inflationary conditions, the lack of minimum
effectively kills the model as there will be no reheating \cite{kofman1} in the classical sense.
(As mentioned previously, gravitational reheating is
far too weak in these models to account for the particle abundance we see today.) 
It is possible to obtain reheating if the tachyon is coupled to several gauge fields \cite{cline}, and is a direction that 
certainly needs further consideration. It is also certainly possible that the potential vanishes for finite $T$,
leading to small oscillations about the minimum \cite{fairbairn} which could provide a mechanism for reheating.
In any case, the issue does not seem to be resolved in a satisfactory manner.

Our geometrical tachyon is no exception to these criticisms. Although the minimum is not at infinity, the effective
theory breaks down when the tachyon rolls to its maximum value and we are unable to proceed. In the 10D gravitational picture
this is due to the probe brane hitting the ring of smeared fivebranes. However even with the simple
form of the DBI action in this instance, we see that outside the ring the potential is approximately exponential \cite{thomas}
and it is suggestive that it may somehow smoothly map onto the cosine at $\rho=R$. One may well enquire what happens if
we consider a case where the fivebranes are not smeared around the ring, rather that they appear resolved 
thus allowing a probe brane
to pass between them.
In this case, we would not expect the effective DBI action to break down and we can
find corrections to the truncated cosine potential and thus obtain
a minimum. This is exactly what we found following the numerical analysis in section 4. Let us now see how the existence of such a minimum can be seen analytically. In order to proceed, we  refer the reader back to the full harmonic function describing $k$ branes at
arbitrary points on the circle \cite{sfetsos}, with the interbrane distance, $x$, given by
\begin{equation}
x = \frac{2\pi R}{k}.
\end{equation}
The full form of the function in the throat region is given by
\begin{equation}
H \sim \frac{kl_s^2}{2R\rho \sinh(y)} \frac{\sinh(ky)}{(\cosh(ky)-\cos(k\theta))},
\end{equation}
where $\rho, \theta$ parameterize the coordinates in the ring plane, and the factor $y$ is given by
\begin{equation}
\cosh(y) = \frac{R^2+\rho^2}{2R\rho}.
\end{equation}
We clearly see that as $k\to \infty$ we recover the expression for the smeared harmonic function which
we used in the previous sections to derive the tachyon potential.
Furthermore we see that when $\rho = R$ the function
reduces to
\begin{equation}
H \sim \frac{k^2l_s^2}{2R^2} \frac{1}{(1-\cos(k\theta))}
\end{equation}
which is clearly finite provided that $\theta \ne 2n\pi/k$, which are the locations of the $NS$5 branes.
In order to
look for a minima we must expand about the point $\rho = R$ using $\rho = R +\xi$, where
$\xi$ is a small parameter which can be
positive or negative. Using the expansion properties of hyperbolic functions
we power expand the harmonic function for an arbitrary fixed angle $\theta$, and we find to leading order\footnote{Thanks to Shinji Tsujikawa for pointing out an algebraic error in a previous draft of this note.}
\begin{equation}
H \sim \frac{k^2l_s^2}{2R^2}\frac{1}{1-\cos(k\theta)} \left(1-|\frac{\xi}{R}|+(5/6- k^2 \frac{2+\cos(k\theta)}{6(1-\cos(k\theta))} )\frac{\xi^2}{R^2}+\ldots \right),
\end{equation}
where we have used the fact that $k\xi << R$ and have neglected any higher order correction terms. Note that the inter brane
distance is given by $2 \pi R /k$ and so our expansion will only be valid for 
distances much smaller than the brane separation. Of course we must be careful not to take
$k$ to be too small since our effective action for the geometrical tachyon will be invalid.
We can clearly see that if the trajectory is at an angle $\theta = (2n+1)\pi/2k$, then the 
harmonic function will reduce to the form (again to leading order in large k)
\begin{equation}
H \sim \frac{k^2l_s^2}{2R^2}\left(1-\frac{k^2\xi^2}{3 R^2}+\ldots  \right)
\end{equation}
We now perform the tachyon map to determine the value of the tachyon as a function
of $\xi$. Note that we expect this 'tachyonic' field to have positive mass squared since it is near the minimum
of its potential. Up to constants we find that
\begin{equation}
T(\xi) \sim \sqrt{\frac{k^2l_s^2}{2(1-\cos(k\theta))}}\left(\frac{\xi}{R} - \frac{\xi^2}{2R^2}+\ldots \right)
\end{equation}
If we assume that the $\xi^2$ term is negligible then we can invert our solution and
calculate the perturbed tachyon potential. Note that keeping higher order terms here does not lead to a simple analytic solution, and so we would hope that
a numerical analysis would be more appropriate. After some manipulation we find 
\begin{eqnarray}
V(T)  &\sim & \frac{\tau_3}{kl_s} \left( 2R^2(1-\cos(k\theta ))\right)^{1/2} 
[ 1+ \frac{T}{2kl_s}\sqrt{2(1-\cos(k\theta))}+  \nonumber \\
 &&T^2 (\, \frac{2+\cos(k\theta )}{6l_s^2} - \frac{(1-\cos(k\theta )\, )}{12k^2l_s^2}\, ) + \ldots ] 
\end{eqnarray}
which shows that the potential is approximately linear around the minimum as it interpolates between the cosine 
and the exponential functions, however this linear term is suppressed by a factor of $1/k$ and we would 
expect it be negligible in the large $k$ limit, thus we can see that there is an approximately quadratic
minimum.
We see that the minimum of the potential in the tachyonic direction will be
\begin{equation}\label{eq:minimum}
V(T(\xi=0))= \frac{\tau_3R}{k l_s}\sqrt{2(1-\cos(k\theta))}
\end{equation}
which can obviously be made small in the large $k$ limit, and will clearly go to zero when the $D3$-brane trajectory
is such that it hits one of the $NS$5-branes. The local maximum will occur at the bisection angle $\theta=\pi/k$, which
we suspect will be an unstable point.
All this fits nicely with our earlier numerical analysis.
Figure 4 in section 4 showed the result of numerical methods used to plot the potential using the exact form of the ring harmonic function. Numerical solutions to the tachyon map inside and outside the ring were matched together to obtain this plot. The minimum can be seen to be quadratic for small distances before mapping onto an exponential function outside
the ring as expected from \cite{thomas}. This is because the numerical analysis includes all the higher order correction terms, which produces a curved potential at the minimum.

The condensing tachyon field may oscillate about the minimum of this potential, assuming that the energy of the 
tachyon is such that it will not overshoot  and return back up the potential toward $T=0$. This 
assumption seems to be valid because as we have just seen the potential no longer has to vanish at $\rho=R$,
so the friction term in (\ref{eq:eom}) will not vanish as the tachyon condenses. However in order for
reheating to occur we must ensure that this term sufficiently damps the motion, confining the field to very
small oscillations about this minimum.

From standard inflationary models we
know that the oscillations about the minimum can be thought of as being a condensate of zero momentum particles
of (mass)$^2$ = V$^{\prime \prime}$(T). The decay of the oscillations leads to the creation of new fields coupled to the
tachyon condensate via the reheating mechanism.
The temperature of this reheating can be approximated as the difference between the maximum and minimum of the potential, and so
we find
\begin{equation}
T_{RH}^4 \sim M_{infl}^4\left(1-\frac{1}{\sqrt{k}}\sqrt{2(1-\cos(k\theta))}\right)
\end{equation}
and so if we assume that the conversion of the tachyon energy is almost perfectly efficient then we will have an upper bound
for the reheating temperature given by the effective inflation scale $M_{infl}$.

We must now consider the more general case where we perturb $\theta$ away from
its bisection value of $\pi/k$. Since we are assuming that the $NS$5-branes are somehow resolvable, we must
also be aware that a single brane does not form an infinite throat \cite{kutasov}. As such, a passing probe brane
will feel the gravitational effect of the fivebranes, but because we expect it to be moving relativistically
we expect that its trajectory will only suffer a slight deflection as it passes by.
In this instance, the perturbed harmonic function at $\rho =R$ reduces to
\begin{equation}
H \sim \frac{k^2l_s^2}{2R^2}\frac{1}{(1+\cos(k\delta))},
\end{equation}
where $\delta$ represents the angular perturbation. Now, we know that
the harmonic function becomes singular when our probe brane hits a five brane so
the function needs to be minimized to ensure a stable trajectory, This is
clearly accomplished by sending $\delta \to 0$. So the value $\pi/k$ corresponds to
an unstable \emph{maximum} from the viewpoint of the tachyon potential.
Of course, we could also see this directly from (\ref{eq:minimum}) by considering perturbations about the bisection
angle.
For small angular deflection we may write
\begin{equation}
H \sim \frac{k^2l_s^2}{4R^2} \left( 1+\frac{k^2 \delta^2}{4}+\ldots \right),
\end{equation}
and so we see that provided $k \delta << 1$ the trajectory of the probe will
be relatively unaffected by the presence of the fivebranes and therefore
we may expect that it will pass between them. On the other hand, for larger
values of $k\delta$, this will not be true and the probe brane may
be pulled into the fivebranes. In any case, we expect that our analysis
of the geometrical tachyon will be invalid in this instance.

The analysis will also be true for a $D3$-brane in a ring $D5$-brane background
using S-duality, the only difference will be to replace
\begin{equation}
kl_s^2 \to 2 g_sk l_s^2,
\end{equation}
where $g_s$ is the string coupling and we again consider $k$ branes on the ring.
The overall effect of switching to the $D5$-brane background is to allow for
a weaker coupling at the top of the potential. In fact the analogue of (\ref{eq:top})
in this picture becomes
\begin{equation}
g_s >> \left(\frac{24 \pi^3 v }{R \sqrt{2k}M_s}\right)^{2/3}
\end{equation}
 The
situation is made slightly more complicated due to the presence of background RR charge,
but this can be neglected when the tachyon is purely time dependent. Thus
we would expect similar results to those obtained in the last two sections. Of course, we should
remember that fundamental strings can end on the $D5$-branes and consequently there can
be additional open string tachyons in the theory.
\section{Discussion}
In this note we have examined the cosmological consequences of the rolling
geometrical tachyon in the early universe. Because of the different mass scale
compared to the open string tachyon, the geometrical tachyon resolves some
of the problems besieging tachyon inflationary models. 

The effective theory is self consistent and appears to be valid description of
4D gravity due to the weak string coupling. This weak coupling arises because
we can select a specific region of our moduli space associated with the
background geometry.
Furthermore we have seen that the form
of the potential satisfies all the slow roll and e-folding conditions, whilst
providing acceptable levels of metric perturbations at the end of inflation. In addition,
we have shown that the potential has a minimum
which will be approximately quadratic for small perturbations in the tachyon field 
and may therefore be used to describe traditional reheating \cite{kofman1}.\footnote{Although
there may be some objections to this due to the non-linearity of the tachyonic action \cite{sami}.}
We have not in any way discussed how reheating can occur in this model, but we have attempted to show that
the potential may well have a metastable minimum which could allow for the kind of field theoretic inflaton oscillation required for standard
reheating.

There are still potential problems associated with this model. Firstly it seems unlikely that we can have an analytic expression for the tachyon valid for any point on the potential due to the complicated nature of the full harmonic function. Thus we have been forced to make approximations or resort to numerical methods. Secondly there is still the issue of fine tuning to deal
with because we need specific values of $k$ and  $ R$ such that a probe brane
passes between fivebranes in the bulk picture without causing the tachyon solution to collapse.
Furthermore we effectively require the $D3$-brane to exhibit one dimensional motion, so that
it continually passes between the $NS$5 branes on the ring as it rolls in the minimum of the potential. More general
motion would imply that the probe brane will eventually hit one of the source branes and cause
the effective theory to break down.
Thirdly we must ensure that the probe brane
is not too energetic, otherwise it may overshoot the minimum and return to the origin. This requires
the friction term to sufficiently damp the motion of the field as it approaches the minimum of the potential.
Although we have argued that this may occur it is not clear whether this requires any fine tuning or not. 
Finally there is the issue of coupling
to other string modes, which will be essential in generating the standard model
fields after reheating, and have been neglected in this and other notes on tachyon inflation.

A more detailed investigation is required before we can rule out this model, 
however we have found some promising results that circumvent
many of the problems associated with tachyon inflation and it is suggestive
that other geometrical tachyon modes may well be viable candidates for the inflaton.
It would also be useful to extend this work to include assisted inflation, and
perhaps non-commutative geometrical tachyons, for example \cite{calcagni}. In addition, it would be useful to
have more understanding of these geometrical tachyon modes and their
relationship to the open string tachyon. In particular the case of $k=2$ is
important \cite{kutasov}, as it relates observables in Little String Theory (LST) to
those in 10D supergravity.
Because the geometrical tachyon fields are dependent upon
the background brane geometry, we may imagine that all these tachyon solutions
correspond to specific points or cross sections of some larger moduli space.
It would therefore be useful to consider other background geometries and how they
are inter-related to some of the known open string tachyon solutions.
We will hopefully return to some of these issues at a future date.

\begin{center} 
\textbf{Acknowledgements}
\end{center}
Many thanks to D. J. Mulryne and N. Pidokrajt for their useful comments and observations.
JW is supported by a QMUL studentship.
This work was in part supported by the EC Marie Curie Research Training Network
MRTN-CT-2004-512194.



\begin{thebibliography}{99}
\bibitem{carroll}
S. M. Carroll, 'TASI Lectures: Cosmology for String Theorists', hep-th/0011110.\\
F. Quevedo, 'Lectures on String/Brane Cosmology', hep-th/0210292.\\
U. H. Danielsson, 'Lectures on string theory and cosmology', hep-th/0409274.
\bibitem{jones}
N. Jones, H. Stioca and S. H. Tye, 'Brane interaction as the origin of inflation', hep-th/0203163.
\bibitem{kofman1}
L. Kofman, A. Linde and A. A. Starobinsky, 'Reheating After Inflation', hep-th/9405187.
\bibitem{kofman2}
L. Kofman and A. Linde, 'Problems with Tachyon Inflation', hep-th/0205121.\\
A. Frolov, L. Kofman, A. Starobinsky, 'Prospects and Problems of Tachyon Matter Cosmology', hep-th/0204187.
\bibitem{linde}
A. Linde, 'Inflation and String Cosmology', hep-th/0503195.
\bibitem{steer}
D. A. Steer and F. Vernizzi, 'Testing and comparing tachyon inflation to single standard field inflation', astro-ph/043290.
\bibitem{freese}
K. Freese and W. H. Kinney, 'On: Natural Inflation', hep-ph/0404012. \\
F. C. Adams, J. R. Bont, K. Freese, J. A. Frieman and A. V. Olinto, 'Natural Inflation:
Particle Physics Models Power Law Spectra For Large-Scale Structure And Constraints From
COBE', hep-ph/9207245.\\
K. Freese, J. A. Frieman and A. V. Olinto, 'Natural Inflation', Phys. Rev. Lett. 65, \textbf{3233} (1990).
\bibitem{sen}
A. Sen, 'Rolling Tachyon', hep-th/0203211, JHEP \textbf{0204} (2002) 048.\\
A. Sen, 'Tachyon Matter', hep-th/0203265.
\bibitem{sen2}
A. Sen, 'Tachyon Dynamics in Open String Theory', hep-th/0410103.
\bibitem{gibbons}
G. W. Gibbons and D. Wiltshire, 'Spacetime as a Membrane in Higher Dimensions', hep-th/0109093, Nucl Phys 
\textbf{B 287} (1987) 717.\\
G. W. Gibbons, 'Cosmological Evolution of the Rolling Tachyon', hep-th/0204008, Phys. Lett. B537 (2002) 1-4.\\
G. W. Gibbons, 'Thoughts on Tachyon Cosmology', hep-th/0301117, Class. Quant. Grav 20 (2003) S321-S346.
\bibitem{cline}
J. M. Cline, H. Firouzjahi, P. Martineau, 'Reheating from Tachyon Condensation', hep-th/0207156.\\
N. Barnaby and J. M.~Cline, 'Creating the universe from brane-antibrane annihilation', hep-th/0403223,
   Phys.\ Rev.\ D {\bf 70}, 023506 (2004).\\
N. Barnaby, C. P. Burgess and J. M. Cline, 'Warped reheating in brane-antibrane inflation', hep-th/0412040.
\bibitem{piao}
Y-S. Piao, R-G Cai, X Zhang and Y-Z Zhang, 'Assisted Tachyonic Inflation', hep-ph/0207143.
\bibitem{joris}
J. Raeymaekers, 'Tachyonic Inflation in a Warped String Background', hep-th/0406195.
\bibitem{li}
X-Z Li, D-J Liu and J-G Hao, 'On the tachyon inflation', hep-th/0207146, J. Shanghai Normal Univ.
(Natural Sciences), Vol.33(4) (2004), 29.
\bibitem{fairbairn}
M. Fairbairn and M. H. G. Tytgat, 'Inflation from a Tachyon Fluid', hep-th/0204070.
\bibitem{shiu}
G. Shiu and I. Wasserman, 'Cosmological constraints on tachyon matter', hep-th/0205003.
\bibitem{tachyon_action}
A. Sen, 'Supersymmetric world-volume action for non-BPS D-branes', hep-th/9909062, JHEP \textbf{9910} (1999) 008.\\
J. Kluson, 'Proposal for non-BPS D-brane action', hep-th/0004106, Phys. Rev. D \textbf{62} (2000) 126003.\\
E. A. Bergshoeff, M. de Roo, T. C. de Witt, E. Eyras and S. Panda, 'T-duality and actions for non-BPS
D-branes', hep-th/0003221, Nucl. Phys. B \textbf{584} (2000) 284.\\
M. R. Garoussi, 'Tachyon couplings on Non-BPS D-branes and Dirac-Born-Infeld action', hep-th/0003122, Nucl. Phys. B \textbf{584} (2000), 284-299.
\bibitem{choudhury}
D. Choudhury, D. Ghosal, D. P. Jatkar and S. Panda, 'On the Cosmological Relevance of the Tachyon',
hep-th/0204204.
\bibitem{time_dependence}
D. Kutasov, 'D-brane Dynamics Near NS5 branes', hep-th/0405058.\\
S. Thomas and J. Ward, 'D-Brane Dynamics and NS5 Rings', hep-th/0411130, JHEP \textbf{0502} (2005) 015. \\
Y. Nakayama, K. Panigrahi, S. J. Rey and H. Takayanagi, 'Rolling Down the Throat
in NS5-Brane Background: The case of Electrified D-Brane', hep-th/0412038, JHEP \textbf{0501} (2005) 052.\\
B. Chen, M. Li and B. Sun, 'Note on DBI dynamics of Dbrane Near NS5-branes', hep-th/0501176.\\
H. Yavartanoo, 'Cosmological solution from D-brane motion in NS5-branes background', hep-th/0407079.\\
A. Ghodsi and A. E. Mosaffa, 'D-brane Dynamics in RR deformation of NS5-branes Background
and Tachyon Cosmology', hep-th/0408015.\\
J. Kluson, 'Non-BPS D-Brane Near NS5-Branes', hep-th/0409298, JHEP \textbf{0411} (2004) 013.\\
J. Kluson, 'Non-BPS Dp-Brane in the background of NS5-Branes on Transverse \textbf{$R^3$}$\times$ \textbf{$S^1$},
hep-th/0411014.\\
K. L. Panigrahi, 'D-Branes Dynamics in Dp-Brane Background', hep-th/0407134, Phys. Lett. B601 (2004) 64-72.
\bibitem{kutasov}
D. Kutasov, 'A Geometric Interpretation of the Open String Tachyon', hep-th/0408073.
\bibitem{thomas}
S. Thomas and J. Ward, 'Geometrical Tachyon Kinks and NS5-Branes', hep-th/0502228.
\bibitem{sahakyan}
D. Sahakyan, 'Comments on D-brane Dynamics Near NS5-Branes', hep-th/0408070. \\
Y. Nakayama, Y. Sugawara and H. Takayanagi, 'Boundary states for the rolling D-branes in NS5-background',
hep-th/0406173, JHEP \textbf{0407} (2004) 020.
\bibitem{bento}
M. C. Bento, O. Bertolami and A. A. Sen, 'Tachyonic Inflation in the Braneworld Scenario', hep-th/0208124.
\bibitem{CHS}
C. G. Callan, J. A. Harvey and A. Strominger, 'Supersymmetric String Solitons', hep-th/9112030.
\bibitem{sfetsos}
Konstadinos Sfetsos, 'Branes for Higgs Phases and Exact Conformal Field Theories', hep-th/9811167, 
JHEP \textbf{9901} (1999) 015.\\
Konstadinos Sfetsos, 'Rotating NS5-brane solution and its exact string theoretic description',
hep-th/9903201, Fortsch. Phy, \textbf{48}, 199 (2000).
\bibitem{hwang}
J. Hwang and H. Noh, 'Cosmological perturbations in a generalized gravity including tachyonic
condensation', hep-th/0206100.
\bibitem{parameters}
J. E. Lidsey, A. R. Liddle, E. W. Kolb, E. J. Copeland, T. Barreiro and M. Abney, 'Reconstructing the Inflaton Potential
- an Overview', astro-ph/9508078.\\
G. Calcagni, 'Slow roll parameters in braneworld cosmologies', hep-th/0402126. 
Phys. Rev. D69, 103508 (2004).
\bibitem{calcagni}
G. Calcagni and S. Tsujikawa, 'Observational constraints on patch inflation in noncommutative spacetime', astro-ph/0407543,  Phys. Rev. D 70, 103514 (2004).
\bibitem{sami}
M. Sami, P. Chingangbam and T. Qureshi, 'Aspects of Tachyonic Inflation with Exponential Potential', hep-th/0205179, Phys.Rev. D66 (2002) 043530.\\
M. R. Garousi, M. Sami and S. Tsujikawa, 'Cosmology from Rolling Massive Scalar Field on the anti-D3 Brane of de Sitter Vacua', hep-th0402075, Phys.Rev. D70 (2004) 043536.








\end{thebibliography}
\end{document}